\documentclass[twocolumn,aps,psfig,epsfig,bm,showpacs,superscriptaddress]{revtex4} 

\usepackage{epsfig}

\begin{document} 

\newcommand{\vk}{{\vec k}} 
\newcommand{\vK}{{\vec K}}  
\newcommand{\vb}{{\vec b}}  
\newcommand{\vp}{{\vec p}}  
\newcommand{\vq}{{\vec q}}  
\newcommand{\vQ}{{\vec Q}} 
\newcommand{\vx}{{\vec x}} 
\newcommand{\vh}{{\hat{v}}} 
\newcommand{\tr}{{{\rm Tr}}}  
\newcommand{\be}{\begin{equation}} 
\newcommand{\ee}{\end{equation}}  
\newcommand{\half}{{\textstyle\frac{1}{2}}}  
\newcommand{\gton}{\stackrel{>}{\sim}} 
\newcommand{\lton}{\mathrel{\lower.9ex \hbox{$\stackrel{\displaystyle 
<}{\sim}$}}}  
\newcommand{\ben}{\begin{enumerate}}  
\newcommand{\een}{\end{enumerate}} 
\newcommand{\bit}{\begin{itemize}}  
\newcommand{\eit}{\end{itemize}} 
\newcommand{\bc}{\begin{center}}  
\newcommand{\ec}{\end{center}} 
\newcommand{\bea}{\begin{eqnarray}}  
\newcommand{\eea}{\end{eqnarray}}

\title{Open Charm and Beauty at Ultrarelativistic Heavy Ion Colliders}
 
\date{\today}
 
\author{Magdalena Djordjevic}
\affiliation{Department of Physics, Columbia University, 
             538 West 120-th Street, New York, NY 10027}
\author{Miklos Gyulassy}
\affiliation{Department of Physics, Columbia University, 
             538 West 120-th Street, New York, NY 10027}
\author{Simon Wicks}
\affiliation{Department of Physics, Columbia University, 
             538 West 120-th Street, New York, NY 10027}

\begin{abstract} 
Important goals of RHIC and LHC experiments with ion beams include the 
creation and study of new forms of matter, such as the Quark Gluon Plasma. 
Heavy quark production and attenuation will provide unique tomographic probes 
of that matter. We predict the suppression pattern of open charm and beauty 
in $Au+Au$ collisions at RHIC and LHC energies based on the DGLV formalism of 
radiative energy loss. A cancelation between effects due to the $\sqrt{s}$ 
energy dependence of the high $p_T$ slope and  heavy quark energy loss is 
predicted to lead to surprising similarity of heavy quark suppression at RHIC 
and LHC. 
\end{abstract}

\pacs{12.38.Mh; 24.85.+p; 25.75.-q}

\maketitle 

{\em Introduction.} RHIC and LHC  experiments involving nuclear collisions 
are designed to create and explore new forms of matter, consisting of 
interacting quarks, antiquarks and gluons. One primordial form of matter, 
called the Quark Gluon Plasma (QGP), is believed to have existed only up to a 
microsecond after the ``Big Bang''. If this QGP phase can be created in 
the laboratory, then a wide variety of probes and observables could be used 
to diagnose and map out its physical properties.
 
The striking discoveries~\cite{QM04} at Relativistic Heavy Ion Collider (RHIC)
of strong collective elliptic flow and light quark and gluon jet quenching, 
together with the decisive null control $d+Au$ data, provide strong evidence 
that a  strongly coupled Quark Gluon Plasma (sQGP), is created in central 
$Au+Au$ collisions at $\sqrt{200}$ AGeV with gluon densities 10-100 times 
greater than nuclear matter densities~\cite{Gyulassy:2004zy}. While there has 
been considerable convergence on the theoretical 
interpretation~\cite{RBRC04} of RHIC data, the experimental exploration of 
the sQGP properties beyond the discovery phase has barely 
begun~\cite{WhitePapers}. Future measurements of rare probes such as direct 
photons, leptons, and heavy quarks will help to more fully map out the sQGP 
properties and dynamics. 

Heavy quarks provide important independent observables that can probe the 
opacity and color field fluctuations in the sQGP produced in high energy 
nuclear collisions. In this letter, we present predictions of open charm and 
beauty quark suppression that can be tested at both RHIC and the future LHC 
facilities. Together with the already established light quark and gluon jet 
quenching and collective elliptic flow, a future observation of a reduced 
heavy quark suppression (as compared to the observed pion suppression) could 
strengthen the current case for sQGP formation as well as test the evolving 
theory of jet tomography~\cite{Gyulassy:2003mc}.

The prediction of D and B meson  suppression pattern, in principle, requires 
theoretical control over the interplay between many competing nuclear 
effects~\cite{Vitev:2002pf} that can modify the $p_{\perp}$ hadron 
spectra of heavy quarks. To study the high $p_{\perp}$ ($p_{\perp}>6$ GeV)
heavy quark suppression, we concentrate on the interplay between two most 
important effects, i.e. jet quenching~\cite{Gyulassy:2003mc,Vitev:2002pf} 
and energy dependence of initial pQCD heavy quark $p_{\perp}$ distribution. 
In addition, we explore a range of initial conditions at LHC based on 
extrapolating RHIC data~\cite{PHOBOS_dNgdy} and based on Color Glass 
Condensate effective theory~\cite{CGC_dNgdy}. We note that, for lower 
$p_{\perp}<6$ GeV spectra nonperturbative effects neglected here, for example 
collective hydrodynamic flow, quark coalescence and the strong gluon shadowing 
in the initial CGC state, may become important~\cite{RBRC04}.

{\em Theoretical framework.} To compute the heavy quark meson suppression we 
apply the DGLV generalization~\cite{Djordjevic:2003zk} of the GLV opacity 
expansion~\cite{Gyulassy:2000er} to heavy quarks. We take into account 
multi-gluon fluctuations as in~\cite{GLV_suppress}.
To apply this method, we need to know the following:
1) Initial heavy quark $p_{\perp}$ distribution,
2) Difference between medium and vacuum gluon radiation spectrum and
3) Heavy quark fragmentation functions.

The initial heavy quark $p_{\perp}$ distributions are computed using the MNR 
code~\cite{MNR}. As in the~\cite{Vogt}, we assume the charm mass to be 
$M_{c}=1.2$ GeV and beauty mass $M_{b}=4.75$ GeV. We assume the same 
factorization and renormalization scales as in~\cite{Vogt}. For simplicity, 
we have concentrated only on bare quark distributions 
($<k_{\perp}^2>=0$ GeV$^2$), and the runs were performed by using CTEQ5M 
parton distributions. 

\begin{figure*}[InitDist]
\hspace*{-0.8cm }\epsfig{file=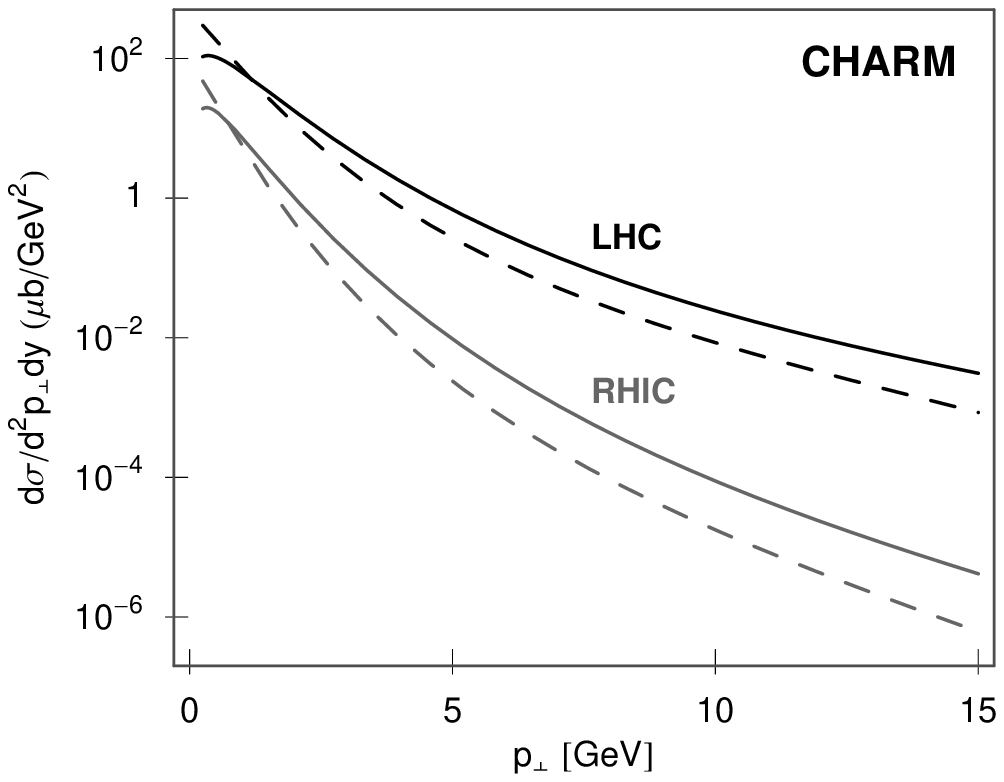,height=2.25in,width=2.8in,clip=5,angle=0} 
\hspace*{0.8cm } \epsfig{file=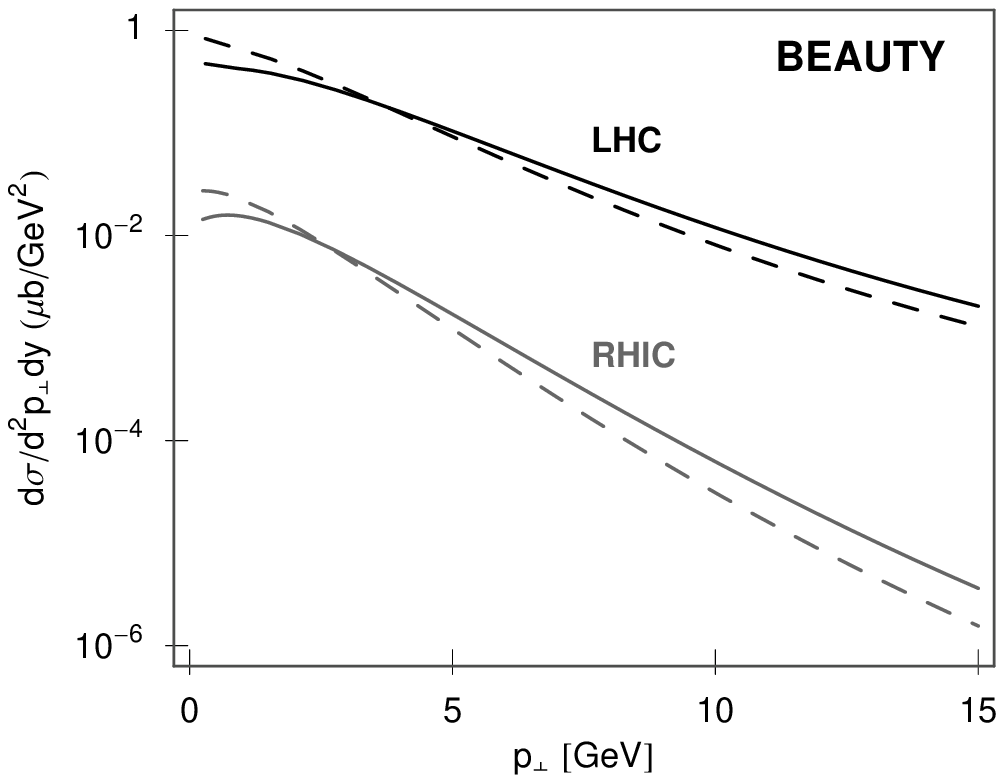,height=2.25in,width=2.8in,clip=5,angle=0} 
\begin{minipage}[t]{17.cm}  
\vspace*{-0.7cm } 
\caption{\label{fig:InitDist} Initial $p_{\perp}$ distributions are shown for 
D (left figure) and B mesons (right figure). Lower (upper) curves correspond 
to RHIC (LHC) case. Solid curves are computed by assuming $\delta$-function 
fragmentation while dashed curves assume Peterson 
fragmentation~\cite{Peterson}. For D (B) mesons we used $\epsilon=0.06$ 
($\epsilon=0.006$)~\cite{Vogt}.}
\end{minipage}
\end{figure*}

Fig.~\ref{fig:InitDist} shows initial $p_{\perp}$ distributions for D and B 
mesons. By comparing $p_{\perp}$ distributions at RHIC and LHC case we see 
that RHIC distributions have significantly larger slope than the LHC ones. 
Since the suppression is sensitive to the slope of quark initial 
$p_{\perp}$ distribution, the decrease in the $p_{\perp}$ slope with the 
increase of collision energy will have the tendency to lower the suppression 
from RHIC to LHC. 

Additionally, comparison between full and dashed curves on 
Fig.~\ref{fig:InitDist} shows the variation of D and B meson $p_{\perp}$ 
distributions using two different types of fragmentation function. Full curves 
show meson spectrum obtained by using $\delta$-function fragmentation, while 
dashed curves show meson spectrum obtained using the Peterson 
fragmentation~\cite{Peterson}. Though the choice of fragmentation function 
can lead to the order of magnitude difference in the absolute $p_{\perp}$, 
we see that slopes of the curves remain quite similar. Therefore, we expect 
that the final suppression is insensitive to the choice of fragmentation 
functions. This conclusion is confirmed in 
Fig.~\ref{fig:dNdyDist} below, 
difference of less than 0.05 in the nuclear modification factor $R_{AA}$ is 
found. ($R_{AA}$ is the ratio of the observed yield in $A+A$ divided by the 
binary collision scaled yield in $p+p$.) Therefore, for clarity, we show most 
results only for $\delta$-function fragmentation for both charm 
and beauty quarks.

To compute the gluon radiation spectrum, we have to include (in general) three 
medium effects that control heavy quark energy loss. These effects are 1)~The 
Ter-Mikayelian, or massive gluon effect~\cite{DG_TM, DG_PLB}, 2)~Transition 
radiation~\cite{Zakharov} which comes from the fact that medium has finite 
size and 3)~Medium induced energy loss~\cite{DG_PLB, Djordjevic:2003zk}, which 
corresponds to the additional gluon radiation induced by the interaction of 
the jet with the medium. 

In~\cite{DG_Trans_Rad} we will show that first two effects are not important 
for heavy quark suppression, since their contribution is less than $10 \%$ in 
the final result. Therefore, in this letter, we address only the medium 
induced gluon radiation spectrum which is given by~\cite{Djordjevic:2003zk}:

\bea 
\frac{ d N_{ind}^{(1)}}{d x} &=&  \frac{C_{F}\alpha_{S}}{\pi} 
\frac{L}{\lambda_g} \int_0^\infty 
\frac{ 2 \mathbf{q}^2 \mu^2 d\mathbf{q}^2}{( \frac{4 E x}{L} )^{2} 
+ (\mathbf{q}^{2} + M^{2}x^{2} + m_{g}^{2})^{2}} 
\nonumber \\
&\times& \int \frac{ d\mathbf{k}^2 \; \theta (2 x (1-x) p_{\perp}-
|\mathbf{k}|)} {(( |\mathbf{k}|-|\mathbf{q}|)^{2} + \mu^{2})^{3/2} 
(( |\mathbf{k}|+|\mathbf{q}|)^{2} + \mu^{2})^{3/2}} 
\nonumber \\
&\times&\left\{ \mu^2+ (\mathbf{k}^{2}-\mathbf{q}^{2}) 
\frac{\mathbf{k}^{2} - M^{2}x^{2} - m_{g}^{2}}
{\mathbf{k}^{2} + M^{2}x^{2} + m_{g}^{2}} \right\}. 
\label{gloun_rad}
\eea

Here, $\mathbf{k}$ is the transverse momentum of the radiated gluon and 
$\mathbf{q}$ is the momentum transfer to the jet. $M$ is heavy quark mass, 
$\mu= 2 (\rho/2)^{1/3}$ is Debye mass, 
$\lambda_g=\frac{8}{9}\frac{\mu^2}{4 \pi \alpha_{S}^2 \rho}$ is mean free 
path~\cite{Gyulassy:2000er}, $m_{g}=\mu / \sqrt{2}$ is gluon mass and 
$E=\sqrt{p_{\perp}^2+M^2}$ is initial heavy quark energy. We assume constant 
$\alpha_{S}=0.3$. For central collisions we take $L=R_x=R_y=6$~fm, and assume 
that $\rho$ is given by (1+1D Bjorken longitudinal expansion~\cite{Bjorken}) 
$\rho={dN_{g}}/{dy \tau \pi L^2} $, where ${\frac{dN_{g}}{dy}}$ is gluon 
rapidity density, and $\tau$ is proper time. 

The energy loss was computed for both 1+1D Bjorken longitudinal expansion and
using an effective {\em average} $\rho$ approximation, where we replace 
$\tau$ by $<\tau>=\frac{L}{2}$. Since both procedures produce similar 
results, in this letter we present only on the computationally simpler 
(average $\rho$) results.

We note that in Eq.(1) 
was set to $k_{max}=2 x (1-x) p_{\perp}$ instead of 
$k_{max}=x E$ used in ~\cite{Djordjevic:2003zk}. Numerically, there is a 
$20 \%$ theoretical uncertainty in $R_{AA}$ due to the different reasonable 
choices of kinematical bounds.

\begin{figure}[ptDist] 
\hspace*{-0.2cm}\epsfig{file=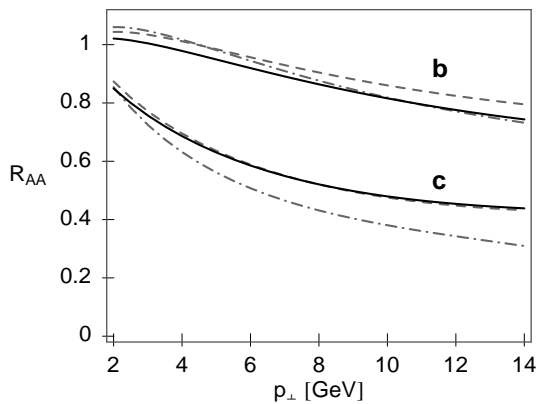,height=2.2in,width=2.8in,clip=5,
angle=0} 
\begin{minipage}[t]{8.6cm}  
\vspace*{-0.7cm} 
\caption{\label{fig:ptDist} The suppression ratio $R_{AA}$ as a function 
of $p_{\perp}$is shown for charm (lower curves) and beauty quarks (upper 
curves). Full curves correspond to RHIC 
case ($\sqrt{s}_{NN}=200$ GeV), while dashed and dot-dashed curves 
correspond to LHC case ($\sqrt{s}_{NN}=5.5$ TeV). Dashed (dot-dashed) 
curves correspond to PHOBOS~\cite{PHOBOS_dNgdy} (CGC~\cite{CGC_dNgdy}) 
extrapolation in gluon rapidity density.}
\end{minipage} 
\end{figure}

{\em Heavy quark suppression at RHIC and LHC.} In this section we compare 
suppression at RHIC and LHC as a function of momentum, collision 
energy and gluon rapidity density dependence. In Fig.~\ref{fig:ptDist} 
we show $R_{AA}(p_{\perp})$ for both charm and beauty quarks corresponding 
to D and B mesons in $\delta$ fragmentation. For estimates of LHC initial 
conditions, we consider  two cases: the PHOBOS 
extrapolation~\cite{PHOBOS_dNgdy} (where gluon density is projected to be 
approximately $60 \%$ higher than at RHIC), and the CGC 
prediction~\cite{CGC_dNgdy} (where the initial gluon density is predicted to be
$\sim 3$ times higher than at RHIC). For charm quark we see that there is 
surprising similarity of $R_{AA}(p_{\perp})$ between RHIC and LHC case, if 
PHOBOS extrapolation in gluon density is assumed. The similarity in 
suppression between these results comes from the fact that, at LHC, the 
enhancement in energy loss (due to the larger gluon density), is mostly 
compensated by the decrease of the heavy quark distribution slopes. A 
slightly greater suppression is obtained with CGC estimate of the initial 
gluon density, which leads to larger energy loss.

By comparing the charm and beauty suppressions on Fig.~\ref{fig:ptDist}, we 
see that significantly less suppression is expected for beauty than 
for charm quarks. This is due to the following two reasons 1) from 
Fig.~\ref{fig:InitDist} we see that beauty $p_T$ distributions have 
significantly smaller slopes than the charm ones, and 2) due to dead cone 
effect~\cite{Dead-done}, the beauty energy loss is much smaller than charm 
energy loss, as shown on Figs.1 and 5 in~\cite{Djordjevic:2003zk}. This 
explains in large part why no significant suppression was observed for 
$p_{\perp}>2$ GeV single electrons at RHIC~\cite{PHENIX_sgle}. In this 
kinematic range there is significant beauty contribution to the single 
electron yields and  that component is essentially unquenched. Cronin and 
possibly collective flow effects in this low $p_{\perp}<6$ GeV region also 
may play a role.

According to Fig.~\ref{fig:ptDist}, we expect similar results for single 
electron suppression at both RHIC and LHC, i.e. we predict no significant 
suppression of single electrons at moderate $p_T$ at LHC as well.

\begin{figure*}[sDist]
\hspace*{-0.3cm }\epsfig{file=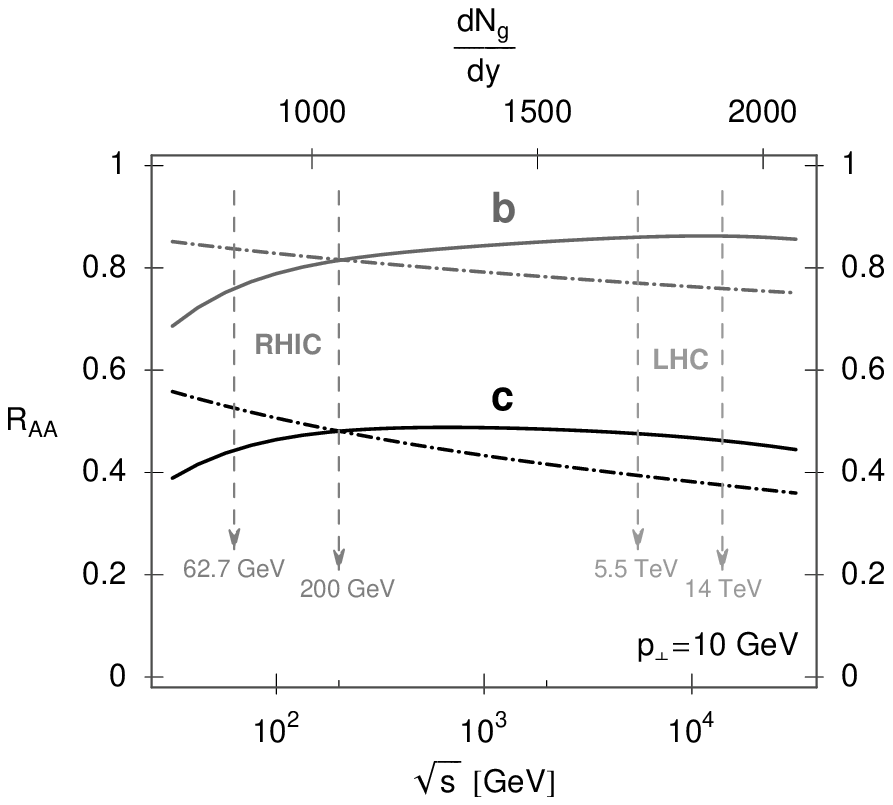,height=2.6in,width=3.2in,clip=5,angle=0} 
\epsfig{file=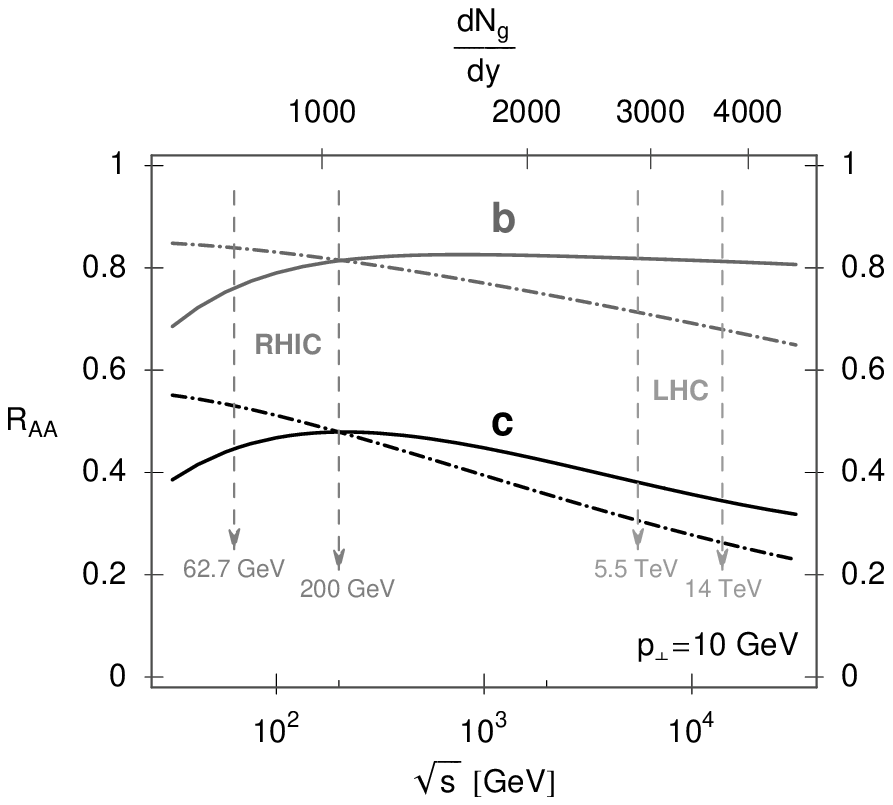,height=2.6in,width=3.2in,clip=5,angle=0} 
\begin{minipage}[t]{17.cm}  
\vspace*{-0.7cm } 
\caption{\label{fig:sDist} The suppression ratio $R_{AA}$ as a function 
of $\sqrt(s)$ is shown for $10$ GeV charm (lower curves) and beauty 
quarks (upper curves). Left (right) figure corresponds to the PHOBOS (CGC) 
extrapolation in gluon rapidity density. Upper x-axis show the gluon rapidity 
density that corresponds to the $\sqrt(s)$ for both PHOBOS and CGC scenario. 
Full curves represent the case where both energy loss and initial quark 
$p_{\perp}$ distribution change with $\sqrt(s)$. Dot-dashed curves correspond 
to the case where only energy loss is changing with $\sqrt(s)$, while initial 
quark $p_{\perp}$ distribution is fixed at $200$ GeV.}
\end{minipage}
\end{figure*}

Our next goal is to study how the suppression is changing as a function of 
collision energy. For that purpose we fix the $p_{\perp}$ of quark jet to 
$10$ GeV and look at $R_{AA}(\sqrt{s})$ as shown on Fig.~\ref{fig:sDist}. 
We see that, if gluon density extrapolates according to PHOBOS, than the 
$RHIC \approx LHC$ conclusion from Fig.~\ref{fig:ptDist} is not a 
coincidence. It rather seams that, in this case, the high $p_{\perp}$ charm 
quark suppression is essentially independent on the collision energy. 
In addition, the slight beauty suppression decreases as the collision energy 
increases. Therefore, we see that in PHOBOS extrapolation case, the $60\%$ 
increase of the gluon density (and equivalently the increase in the energy 
loss) is not enough to compensate the decrease in the $p_{\perp}$ slope. 

Slightly different situation occurs in the case of CGC extrapolation in gluon
density. In this case, at LHC, we can expect $20 \%$ higher suppression for 
charm quarks and constant suppression for beauty quarks. 

Therefore, the main conclusion following from Figs.~\ref{fig:ptDist} 
and~\ref{fig:sDist}, is that no significant difference between the RHIC and 
LHC heavy quark suppression is expected. This result is surprising. To 
emphasize this point, we show in Fig.~\ref{fig:sDist} dot-dashed curves 
showing a hypothetical case in which we assume that only energy loss changes 
with collision energy, while heavy quark initial $p_{\perp}$ distribution 
remains unchanged and fixed to $200$ GeV case. From this curves we see that, 
at LHC, the energy loss would lead to additional $0.1$ decrease in $R_{AA}$ 
for both PHOBOS and CGC case. 

\begin{figure}[dNdyDist] 
\hspace*{-0.2cm}\epsfig{file=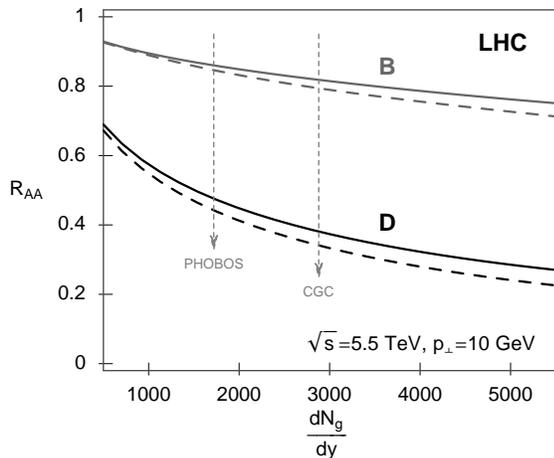,height=2.4in,width=3.in,clip=5,
angle=0} 
\begin{minipage}[t]{8.6cm}  
\vspace*{-0.7cm} 
\caption{\label{fig:dNdyDist} The suppression ratio $R_{AA}$ as a function 
of gluon density is shown for D (lower curves) and B mesons (upper curves).
Solid curves are computed by assuming $\delta$-function 
fragmentation while dashed curves assume Peterson 
fragmentation~\cite{Peterson}. For D (B) mesons we used $\epsilon=0.06$ 
($\epsilon=0.006$)~\cite{Vogt}.}
\end{minipage} 
\end{figure} 
 
If we compare the suppression for PHOBOS and CGC case on 
Fig.~\ref{fig:sDist}, we see that at $5.5$ TeV (LHC) 
the difference of $1000$ in gluon rapidity density leads to only $\approx 0.1$ 
difference in $R_{AA}$. Since the $\frac{dN_{g}}{dy}$ is still unknown at LHC,
on Fig.~\ref{fig:dNdyDist} we show $R_{AA} (\frac{dN_{g}}{dy})$ for a $10$ 
GeV D and B mesons. We see that both D and B meson suppression falls 
slowly with the increase of the initial gluon rapidity density.

{\em Conclusions.} In this letter we predicted the nuclear modification factor
$R_{AA}(p_T, M_Q,\sqrt{s}, \frac{dN_{g}}{dy})$ for charm and beauty quark 
production in central $Au+Au$ reactions with $\sqrt{s}=200-5500$ AGeV.  
We predict a rather weak $\sqrt{s}$ dependence in this range due to the 
compensation of the increasing energy loss in the more opaque sQGP and 
kinematic reduction of the $p_T$ slope. Of course, it is still 
straightforward to deconvolute these competing effects to determine the growth 
of initial density with $\sqrt{s}$ and therefore differentiate between 
different predictions, such as CGC, of those initial conditions.

By comparing our heavy quark predictions to the suppression patterns for the 
neutral pions in Ref.~\cite{Vitev:2002pf} (light quark and gluon case), we 
expect a striking difference in the suppression pattern between light and 
heavy mesons. This is because the much more strongly quenched gluon jets 
component of light hadrons does not play a role in D and B production.
The light hadron quenching pattern is therefore expected to have a stronger 
collision energy dependence~\cite{Vitev:2002pf}.

We expect a moderate D meson suppression $R_{AA}\approx 0.5 \pm 0.1$ for the 
$\frac{dN_g}{dy} \approx 1000\pm 200$ inferred from $\pi^0$. A similar 
suppression is expected at LHC for $1.5-3$ times larger $\frac{dN_g}{dy}$. 
Our high $p_{\perp}>6$ GeV predictions are robust within our approach, and 
significant experimental deviations would pose serious challenge to the pQCD 
based theory of radiative energy loss in sQGP matter. Future $D$ meson data 
on 200 GeV $d+Au$ and $Au+Au$ and eventually at LHC will thus enable critical 
consistency tests of the theory and the tomographic inferences drawn from the 
observed jet quenching patterns.

{\em Acknowledgments:} Authors would like to thank R. Vogt on her help with 
MNR code, and valuable discussions about the heavy quark production. This 
work is supported by the Director, Office of Science, Office of High Energy 
and Nuclear Physics, Division of Nuclear Physics, of the U.S. Department of 
Energy under Grant No. DE-FG02-93ER40764.

\end{document}